\documentclass[]{spie}  

\usepackage{amsmath,amsfonts,amssymb,xspace}
\usepackage{graphicx}
\usepackage[colorlinks=true, allcolors=blue]{hyperref}

\newcommand{\mum}{\mbox{{\usefont{U}{eur}{m}{n}{\char22}}m}\xspace}
\newcommand{\mus}{\mbox{{\usefont{U}{eur}{m}{n}{\char22}}s}\xspace}

\title{AO3000 at Subaru: Combining for the first time a NIR WFS using First Light’s C-RED ONE and ALPAO’s 64x64 DM}

\author[a]{Julien Lozi}
\author[a]{Kyohoon Ahn}
\author[b]{Christophe Clergeon}
\author[a]{Vincent Deo}
\author[a,c,d,e]{Olivier Guyon}
\author[a]{Takashi Hattori}
\author[a]{Yosuke Minowa}
\author[f]{Shogo Nishiyama}
\author[a]{Yoshito Ono}
\author[a,e,g]{S\'{e}bastien Vievard}
\affil[a]{Subaru Telescope, National Astronomical Observatory of Japan, National Institutes of Natural Sciences (NINS), 650 North A`oh\={o}k\={u} Place, Hilo, HI 96720, United States}
\affil[b]{Gemini International Observatory, a program of NSF's NOIRlab, 670 North A`oh\={o}k\={u} Place, Hilo, HI 96720, United States}
\affil[c]{Steward Observatory, University of Arizona, Tucson, AZ 87521, United States}
\affil[d]{College of Optical Sciences, University of Arizona, Tucson, AZ 87521, United States}
\affil[e]{Astrobiology Center of NINS, 2 Chome-21-1, Osawa, Mitaka, Tokyo, 181-8588, Japan}
\affil[f]{Miyagi University of Education, 149, Aramaki-aza-Aoba, Aobaku, Sendai}
\affil[g]{Observatoire de Paris, LESIA, 5 Place Jules Janssen, 92190 Meudon, France}

\authorinfo{Further author information: (Send correspondence to J.L.)\\J.L.: E-mail: lozi@naoj.org, Telephone: 1 808 934 5949}

\begin{document} 
\maketitle

\begin{abstract}
After 16 years of on-sky operation, Subaru Telescope’s facility adaptive optics AO188 is getting several major upgrades to become the extreme-AO AO3000 (3000 actuators in the pupil compared to 188 previously). AO3000 will provide high-Strehl images for several instruments from visible to mid-infrared, notably the Infrared Camera and Spectrograph (IRCS), and the Subaru Coronagraphic Extreme Adaptive Optics (SCExAO). For this upgrade, the original 188-element deformable mirror (DM) will be replaced with ALPAO’s $64\times64$ DM. The visible wavefront sensor will also be upgraded at a later date, but in the meantime we are adding a near-infrared Wavefront Sensor (NIR WFS), using either a double roof prism pyramid mode or a focal plane WFS mode. This new wavefront sensor will use for the first time First Light’s C-RED ONE camera, allowing for a full control of the $64\times64$ DM at up to 1.6~kHz. One of the challenges is the use of non-destructive reads and a rolling shutter with the modulated pyramid. This upgrade will be particularly exciting for SCExAO, since the extreme-AO loop will focus more on creating high-contrast dark zones instead of correcting large atmospheric residuals. It will be the first time two extreme-AO loops will be combined on the same telescope. Finally, the setup AO3000+SCExAO+IRCS will serve as a perfect demonstrator for the Thirty Meter Telescope’s Planetary Systems Imager (TMT-PSI). We will present here the design, integration and testing of AO3000, and show the first on-sky results.
 
\end{abstract}


\section{INTRODUCTION}
\label{sec:intro}  

The first generations of adaptive optics (AO) and extreme AO (XAO) are getting upgraded with a larger number of actuators, faster low-noise detectors, more sensitive wavefront sensors (WFS) and even new near-infrared WFS (NIR WFS).

At the Subaru Telescope, following the success of the first AO system AO36, AO188 was developed and installed in 2006\cite{Takami2006}, including a laser guide star\cite{Hayano2006}. With 188 actuators in the pupil, AO188 can provide Strehl ratios of $\sim20$--40\% in H-band in median seeing\cite{Minowa2010}. Three instruments are using the output of AO188:
\begin{itemize}
    \item The facility instrument IRCS (Infrared Camera and Spectrograph), providing both imaging (y- to M'-bands) and spectroscopy (zJ- to L-band) capabilities.
    \item The PI XAO platform SCExAO (Subaru Coronagraphic Extreme Adaptive Optics), feeding several science modules in visible (R- and i-band) and NIR (y- to K-band), including the integral field spectrograph CHARIS, the MKID Exoplanet Camera (MEC) and the visible differential polarimetric imager VAMPIRES\cite{Jovanovic2015,Lozi2018}.
    \item the PI instrument IRD (Infrared Doppler spectrograph), a NIR (y- to H-band) fiber-fed high resolution spectrograph. IRD can be fed directly with multi-mode fibers from AO188, or with single-mode fibers from SCExAO (REACH module)\cite{Kotani2018}.
\end{itemize}

In the broader context of Subaru 2\footnote{Subaru 2 website: \url{https://subarutelescope.org/jp/subaru2/}}, aiming at enhancing Subaru's functionalities in key aspects, such as wide-field, high-resolution observations and infrared astronomy, we want to push for better and faster wavefront control by upgrading components of AO188, SCExAO, and the whole instrument architecture of the infrared Nasmyth platform of Subaru. Finally, we want to test technologies necessary to prepare for the next generation of giant segmented telescopes like the Thirty Meter Telescope (TMT)\cite{Ono2020}.

In this paper, we will first present the steps envisioned to upgrade the facility adaptive optics and the instrument configuration (Sec.~\ref{sec:phases}). Then we will detail the work performed on a new NIR WFS (Sec.~\ref{sec:nirwfs}), including the design and integration, and the first on-sky testing of its modes. Then we will present the preliminary steps taken to upgrade the DM (Sec.~\ref{sec:dm}). Finally, we will conclude and give some perspective to the work ahead of us.

\section{AO3000: AN UPGRADE OF AO188 IN PHASES}
\label{sec:phases}

The current facility adaptive optics of the Subaru Telescope AO188 is composed of a 188-actuator bimorph DM from Cilas, coupled with a visible curvature wavefront sensor using avalanche photo-diodes (APDs). A vibrating membrane modulate the signal at 2~kHz, So the loop runs at 1~kHz. It provides Strehl ratios between 20 and 40\% in H-band in median seeing. 

Higher performances are now becoming essential to improve the scientific output of the instruments using AO, by reaching deeper contrasts for high-contrast imagers, reaching fainter targets for imaging and spectroscopy, and reaching redder targets (M-type stars, stars embedded in dust, e.g. with protoplanetary rings, or around the galactic center)

The goal is to upgrade the facility adaptive optics in phases, with minimal impact on the observing schedule and current functionalities.

   \begin{figure} [tb]
   \begin{center}
   \begin{tabular}{c}
   \includegraphics[width=0.97\textwidth]{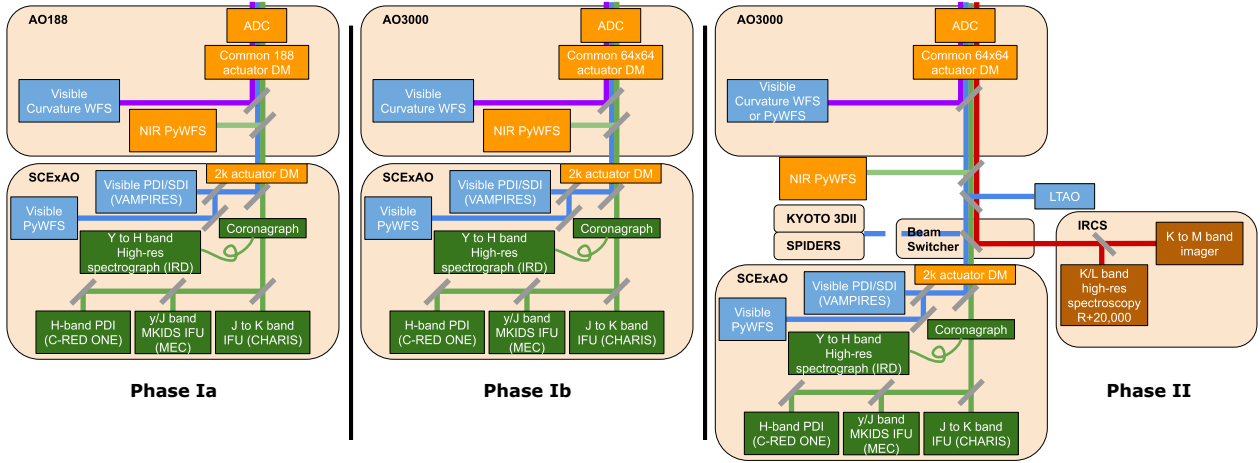}
   \end{tabular}
   \end{center}
   \caption{Phases of upgrades to the facility adaptive optics AO188 to AO3000. Phase Ia: addition of a NIR WFS inside AO188. Phase Ib: Replacement of the DM to ALPAO's 64x64. Phase II: Addition of a NasIR beam switcher to split the light between instruments.}
   \label{fig:phases} 
   \end{figure} 

\begin{itemize}
    \item \textbf{Phase Ia: Addition of a NIR WFS} (see Fig.~\ref{fig:phases} left).
    
    The goal of this phase is to add a NIR WFS inside the existing enclosure, without affecting the current AO188 performances and functionalities. This is the main focus of this paper. It was integrated inside AO188 for the first time in May 2022, and tested on-sky during two engineering half-nights on May 21 and May 24, although due to high humidity, only less than 2~hours of data could be obtained. Due to concerns for the safety of the bimorph DM, it was decided to only take open-loop data, and only perform closed-loop tests in the next phase.
    \item \textbf{Phase Ib: Replacement of the original 188-actuator DM with ALPAO’s 64x64 DM}\cite{Vidal2019}  (see Fig.~\ref{fig:phases} middle).
    
    In this phase, the number of available actuators in the pupil will increase from 188 to $\sim3000$. Therefore, AO1888 will be re-branded as AO3000. The DM replacement is expected to be tested on-sky in closed-loop with the NIR WFS at the end of November 2022, although the priority will be to maintain the current performance of AO188 by down-sampling the DM and using the current APDs for wavefront control.
    
    The new DM is a drop-in replacement of the original one, although the original one is also mounted on a fast tip/tilt mount where offloads of these modes are sent. Therefore, with the new DM, we will have to offload larger tip/tilt commands directly to the secondary mirror of the telescope, although we can only do it at lower speeds. We are also studying the possibility of reusing the fast tip/tilt mount for another flat mirror inside AO188.
    
    The visible curvature WFS will be upgraded with a non-linear version of the sensor using a sCMOS camera and imaging simultaneously four out-of-focus pupil images. This new WFS is still in an experimental phase, and the timeline of its integration is still undecided.
    
    \item \textbf{Phase II: Installation of a NasIR beam switcher for up to 4 instruments}  (see Fig.~\ref{fig:phases} right).
    
    A NasIR beam switcher (NBS) is in its final design phase, allowing to switch rapidly between up to 4 instruments, and even splitting the light between 2 instruments (e.g. SCExAO and IRCS) with a dichroic beamsplitter.
    
    In this phase, the NIR WFS will move out of AO3000, on a common platform with a future laser tomography AO (LTAO), a precursor of the future MCAO ULTIMATE-SUBARU. In this configuration, the NIR WFS will be equipped with a pickoff wheel that will host the various dichroic beamsplitters described in the next section.
\end{itemize}

\section{PHASE Ia: NIR WFS}
\label{sec:nirwfs}

\subsection{NIR WFS: Design and Integration}
\label{sec:design}

A NIR PyWFS was first envisioned in the development phase of AO188, based on the PYRAMIR system\cite{Feldt2006a,Feldt2006b}, but it did not materialize. It was not until the development of the photon-counting NIR SAPHIRA detector\cite{Atkinson2018} that a NIR PyWFS was tested on the SCExAO instrument as a technology demonstration\cite{Lozi2019}.

\begin{figure} [b]
   \begin{center}
   \begin{tabular}{c}
   \includegraphics[width=0.8\textwidth]{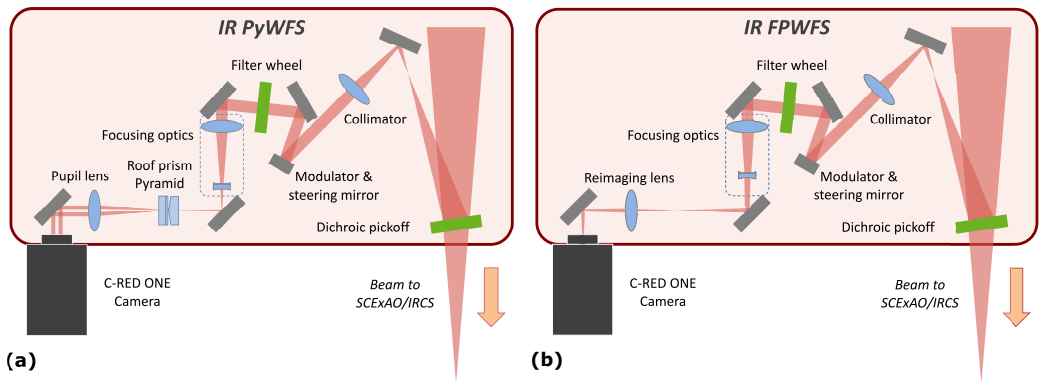}
   \end{tabular}
   \end{center}
   \caption{Diagram of the optical path of the NIR WFS inside AO18. (a) PyWFS mode. (b) FPWFS mode. A pupil viewing mode is also available by using the PyWFS mode and moving the pyramid optics out of the way.}
   \label{fig:diagram} 
   \end{figure}

The opto-mechanical design AO3000's NIR WFS was based on the legacy of SCExAO's visible pyramid WFS (PyWFS): the light from the guide star is collimated and the pupil reimaged on a fast tip/tilt modulator using a custom achromatic doublet ($f=150$~mm, $\phi_\textrm{pupil}=10.7$~mm), then it is refocused on the tip of a double-roof prism pyramid optics, using a pair of converging and diverging doublets ($f_1=150$~mm and $f_2=-50$~mm). Finally, a pupil lens ($f=80$~mm) reimages the four pupils on a fast camera (see Fig.~\ref{fig:diagram} (a)).

With SCExAO, we demonstrated that the double-roof prism design was performing well, despite its chromatic nature. The NIR version is designed to work between y- and H-band (0.98 to 1.8~\mum), while the visible one works between 0.6 and 0.9~\mum. The visible one uses N-BK7 glass, while the NIR one uses CaF2 glass, for its lower chromaticity in infrared. In the end, both designs end up with a similar chromatic spread over their working range.

The camera used for wavefront sensing is the First Light Imaging C-RED ONE camera, a fast low-noise photon-counting device that is currently the most suited for this application. The camera was purchased for two applications: the NIR WFS described here, and a fast polarization differential imaging (FPDI) mode, that was tested and offered for Open-Use science in S21B\cite{Lozi2020}. The opto-mechanical design had to accommodate the fact that the camera would physically need to be moved between two locations, although since then, we managed to purchase a second camera. 

The main difference with SCExAO's PyWFS is that we need to be able to close the loop on off-axis natural guide stars. One application for the NIR WFS is the observation of the galactic center, where the closest guide star (GCIRS 7, $mag_H = 9.26$) is 5" away from Sgr A*. To be able to reach this type of targets, the modulator is mounted on a motorized tip/tilt mount, which allows to steer the line of sight of the NIR WFS. 

Finally, during the design phase, it was decided to add a focal plane imaging mode to the NIR WFS for several reasons:
\begin{itemize}
    \item It allows to calibrate the radius of modulation,
    \item It is useful to measure seeing before closing the loop,
    \item It can characterize the NIR point spread function (PSF) when the visible wavefront sensor is used,
    \item by adding defocus, it ca be used as a focal plane WFS on faint targets, unreachable with the PyWFS mode. This can eventually be useful as a truth WFS (tip/tilt and focus modes for the LTAO.
\end{itemize}
The work on focal plane wavefront sensing is on-going with SCExAO, and is mostly based on the legacy of the Lyot-based Low-Order WFS (LLOWFS)\cite{Singh2015} and mono-plane wavefront sensing\cite{Vievard2020}.

The FPWFS mode is achieved simply by moving the pyramid optics out of the way, changing the pupil lens with a reimaging lens ($f=150$~mm), and refocusing the diverging lens of the focusing optics (see Fig.~\ref{fig:diagram} (b)). Thanks to this modularity, a simple pupil viewing mode is also available by using the PyWFS mode and moving the pyramid optics out of the way.

Mechanically, the difficulty was to have a compact design that uses the limited space inside AO188, without affecting the current components. It uses mainly off-the-shelf parts and motors, and very limited custom parts. 

   \begin{figure} [t]
   \begin{center}
   \begin{tabular}{c}
   \includegraphics[width=0.97\textwidth]{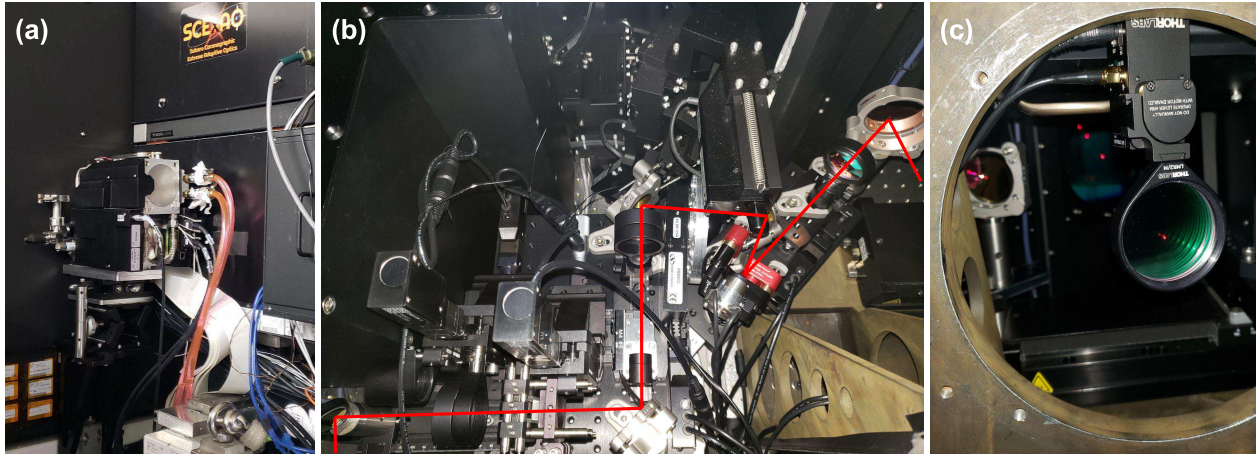}
   \end{tabular}
   \end{center}
   \caption{Integration of the NIR WFS inside AO188. (a) C-RED ONE camera sticking out of AO188, with SCExAO on its right. (b) NIR WFS core optics inside AO188. (c) NIR WFS dichroic beamsplitter in a flip mount, at the output port of AO188.}
   \label{fig:nirwfs_pics} 
   \end{figure} 

The design of the NIR WFS was completed in 2021. All the parts were procured and assembled in 2021 --beginning of 2022, and mounted inside AO188 for the first time in May 2022. Figure~\ref{fig:nirwfs_pics} shows the final result: (a) we can see the C-RED ONE camera mounted on a platform sticking out of AO188. On its right is the SCExAO instrument; (b) the core optics of the NIR WFS inside AO188, despite the tight space; (c) the dichroic beamsplitter sending part of the NIR light to the sensor. In this phase, only one beamsplitter can be mounted at a time in the motorized flip mount. 

   \begin{figure} [b]
   \begin{center}
   \begin{tabular}{c}
   \includegraphics[width=0.97\textwidth]{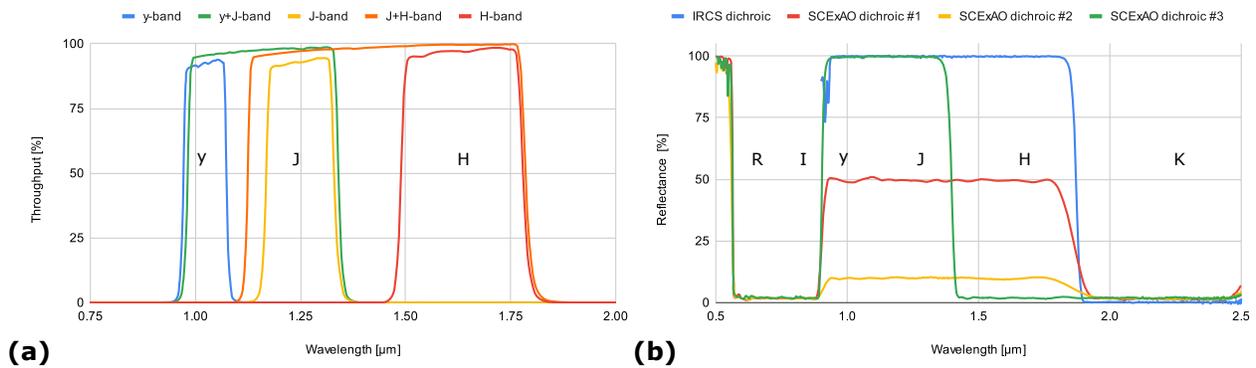}
   \end{tabular}
   \end{center}
   \caption{(a) Throughput of the filters inside the NIR WFS filter wheel. (b) Reflectivity of the NIR WFS dichroic pickoff options available.}
   \label{fig:nirwfs_throughput} 
   \end{figure} 

The wavelength range of the NIR WFS is defined by two optics: the dichroic beamsplitter and an internal filter wheel. The filter wheel contains the regular Maunakea band filters from ASAHI Spectra (y-, J- and H-band), as well as two custom filters combining respectively y- and J-band, and J- and H-band. Their throughputs are compiled in Fig.~\ref{fig:nirwfs_throughput} (a). An open slot is also available to get the whole broadband mode, between y- and H-band.
For the dichroic beamsplitters, we currently have four options. Their reflectance are presented in Fig.~\ref{fig:nirwfs_throughput} (b). Their absorption is negligible, so the rest of the light goes to the science instrument. The four beamsplitters are:
\begin{itemize}
    \item \textbf{IRCS dichroic}: designed for IRCS K-band observations, it sends all of y- to H-band to the NIR WFS, and all of K-band to the science instrument. It can also be used for SCExAO/CHARIS K-band high-resolution IFS.
    \item \textbf{SCExAO dichroic \#1}: It sends 50\% of y- to H-band to the NIR WFS, and the other 50\% to SCExAO. It also transmit all of K-band and all of visible light (R- and I-bands) to the science modules.
    \item \textbf{SCExAO dichroic \#2}: Similarly to \#1, all of R-, I- ad K-band is transmitted, and only 10\% of y- to H-band is reflected towards the NIR WFS. This dichroic would be more suited for brighter targets.
    \item \textbf{SCExAO dichroic \#3}: This beamsplitter sends all of y- and J-band to the NIR WFS, and transmits all of R-, I-, H- and K-band for science.
\end{itemize}
Although each beamsplitter was designed for a specific instrument, they could be used by others if they are more suited. Once the NBS is installed for phase II, the NIR WFS will be able to migrate out of the AO3000 enclosure, which will allow us to use a beamsplitter wheel instead of the flip mount. In that case, all the beamsplitters will be available without manual intervention. We may get other beamsplitter designs as well in the future.

The C-RED ONE camera is used in Correlated Double Sampling (CDS) mode, where the sub-frame is read immediately after a frame reset, then reset and read again. The result is the difference between the frame before the reset and the previous one just after the reset. The sub-frame used is 160x160 pixels, with a read time of $t_\textrm{read}=102$~\mus. So if we program the CDS mode to run at a frequency $F_\textrm{CDS}$, then the actual integration time $t_\textrm{int}$ will be
\begin{equation}
    t_\textrm{int} = \frac{1}{F_\textrm{CDS}} - t_\textrm{read}.
\end{equation}
In that case, we do not want to modulator to be synchronized with the CDS reads, otherwise we would lose a fraction of the rotation during the actual integration time. Therefore, the modulation frequency $F_\textrm{mod}$ is set to
\begin{equation}
    F_\textrm{mod} = \frac{1}{t_\textrm{int}} = \frac{F_\textrm{CDS}}{1-F_\textrm{CDS}t_\textrm{read}}.
\end{equation}

\begin{table}[b]
\caption{Integration time, modulator frequency and maximum modulation radius for various loop frequencies.} 
\label{tab:modulation}
\begin{center}       
\begin{tabular}{|r|c|c|c|c|c|} 
\hline
\rule[-1ex]{0pt}{3.5ex}  CDS frequency [Hz] & 500 & 750 & 1000 & 1500 & 2000  \\
\hline
\rule[-1ex]{0pt}{3.5ex}  Actual integration time [ms] & 1.898 & 1.231 & 0.898 & 0.565 & 0.398  \\
\hline
\rule[-1ex]{0pt}{3.5ex}  Modulator frequency [Hz] & 526.9 & 812.1 & 1113.6 & 1771.0 & 2512.6  \\
\hline
\rule[-1ex]{0pt}{3.5ex}  Duty cycle [\%] & 95 & 92 & 90 & 85 & 80  \\
\hline
\rule[-1ex]{0pt}{3.5ex}  Maximum modulation radius [mas] & 250 & 190 & 145 & 60 & 40  \\
\hline 
\end{tabular}
\end{center}
\end{table}

Table~\ref{tab:modulation} presents the integration time, modulator frequency and maximum modulation radius for various loop frequencies. The maximum frequency for CDS mode and the desired sub-window is 4.9~kHz, at which point the duty cycle drops to 50\%. The readout scheme creates a rolling shutter effect linked to the duty cycle that is out of the scope of this paper, although we will probably always be at or below 2~kHz, where this effect will not impact us too much.

   \begin{figure} [t]
   \begin{center}
   \begin{tabular}{c}
   \includegraphics[width=0.97\textwidth]{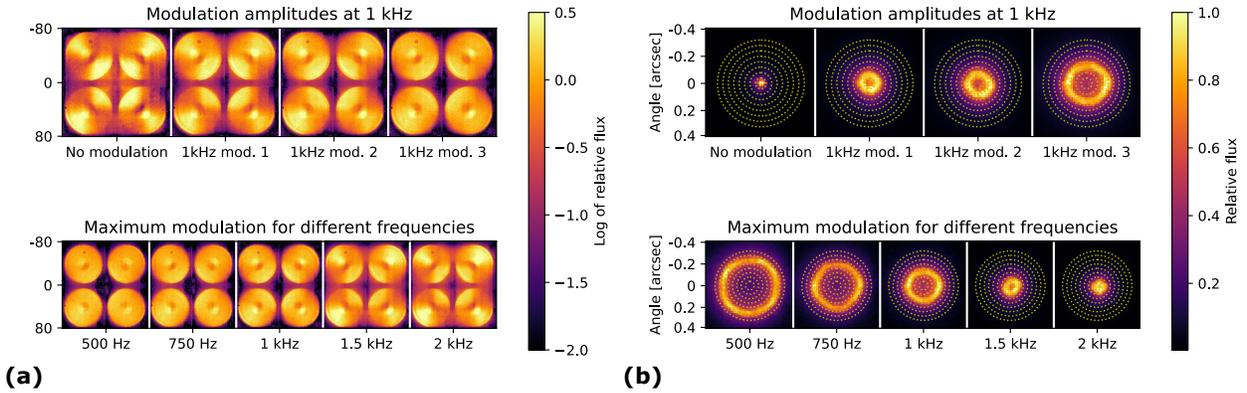}
   \end{tabular}
   \end{center}
   \caption{(a) PyWFS images for different frequencies and modulation using the internal laser source (1550~nm). The “flat” DM shows signs of astigmatism. (b) Different modulation parameters @1 kHz (top) and maximum modulation at different frequencies (bottom).}
   \label{fig:labmodulation} 
   \end{figure}

AO188 is only equipped with two laser diodes, one at 633~nm, and one at 1.55~\mum, but lacks a proper broadband source. We will upgrade the calibration source in the future, but for now, we could only use the NIR diode for internal testing. We programmed modulator parameters for the frequencies mentioned in Tab.~\ref{tab:modulation}, as well as three different amplitudes at 1~kHz. The modulator can always be turned off at any frequency.

Figure~\ref{fig:labmodulation} presents results using the internal laser diode, (a) in PyWFS mode and (b) in focal plane mode. The PyWFS images are displayed in log scale to see the amount of light between the pupils. For the focal plane images, circles with radii steps of 40~mas (1~$\lambda/D$ in H-band) were added to do a quick measurement of the modulation radii.

From Fig.~\ref{fig:labmodulation} (a), we can notice a few things. First, the supposed "flat" map of the bimorph DM is not quite flat, some astigmatism is visible. Second, the actuators behind the central obstruction create an artifact on the wavefront at the center. These two points are probably due to the fact that the source is not perfectly aligned, and there is not pupil mask simulating the Subaru pupil when creating a flat map with the internal source. Besides that, we can see that there is a minimal amount of light between the pupils, even when the modulator is turned off, indicating a very good quality of the roof prism vertices. We can also notice that, as expected, the amount of light outside of the pupils is reduced with increased modulation. Similarly, the contrast of the astigmatism mode is reduced with higher modulation.

On Fig.~\ref{fig:labmodulation} (b), we can see the corresponding modulations in the focal plane. Some work is still needed at some frequencies to get more circular patterns, and also to be able to transition smoothly between modulation radii. In the future, we would like to be able to reduce the modulation on the fly, depending on the seeing and the quality of the correction, and potentially turn off the modulation. 

\subsection{First On-Sky Testing of the PyWFS Mode}
\label{sec:pywfs}
   \begin{figure} [b]
   \begin{center}
   \begin{tabular}{c}
   \includegraphics[width=0.97\textwidth]{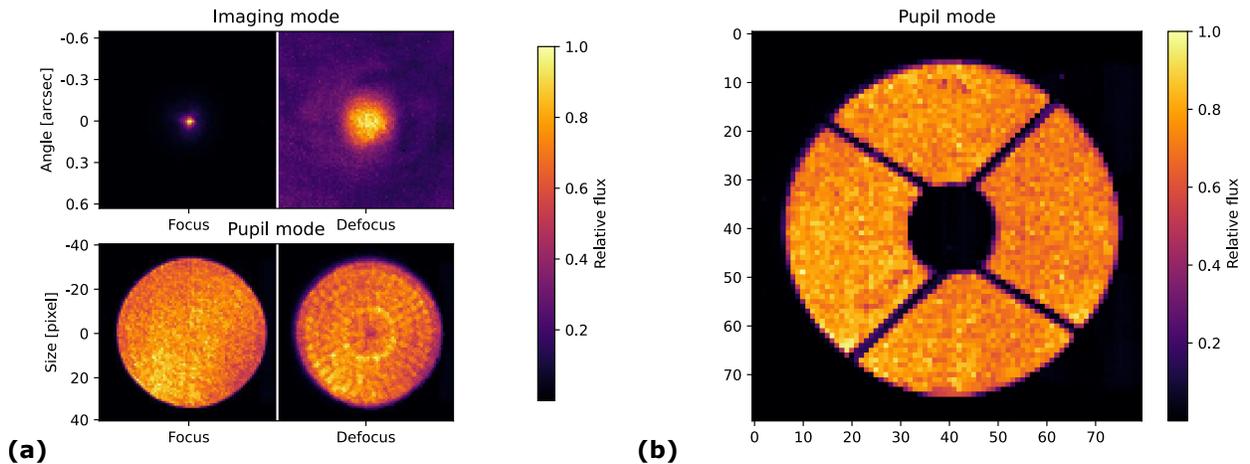}
   \end{tabular}
   \end{center}
   \caption{(a) Focal plane and pupil plane image in and out of focus, using the internal 1550~nm laser diode. (b) Pupil mode on-sky, in broadband mode (y- to H-band).}
   \label{fig:pupil} 
   \end{figure} 
The testing with the lab source were very limited, due to the quality of the source, and the lack of broadband light. Figure~\ref{fig:pupil} (a) present focal plane and pupil images without and with added defocus. The defocused pupil image shows the actuators of the bimorph DM, especially the effect of central actuators creating a ring of aberrations. The defocused focal plane image also shows that the image has some halo , maybe due to the diode itself.

The NIR WFS was tested on-sky for the first time during two engineering nights, on May 21 and May 24, 2022, in median seeing conditions ($\sim0.6$") for both nights. The SCExAO dichroic \#1 (referred as the SCExAO dichroic hereafter) was the one used as the beamsplitter, allowing us the record visible and NIR telemetry with SCExAO in the background. We only managed to get about 2 hours of data due to high humidity. Despite the short amount of time, we acquired sufficient data to characterize all the modes. For safety reasons, no command from the NIR WFS was sent to the original bimorph DM, although the wavefront control loop was still closed using the visible curvature WFS. 

Figure~\ref{fig:pupil} (b) validates the pupil imaging mode: the pupil of Subaru, including the spider arms and central obstruction, are well conjugated. The pupil is about 68~pixels across, as expected. It was designed to over-sample slightly the new ALPAO DM, which should provide 60 to 62~actuators across the pupil diameter.

   \begin{figure} [t]
   \begin{center}
   \begin{tabular}{c}
   \includegraphics[width=0.97\textwidth]{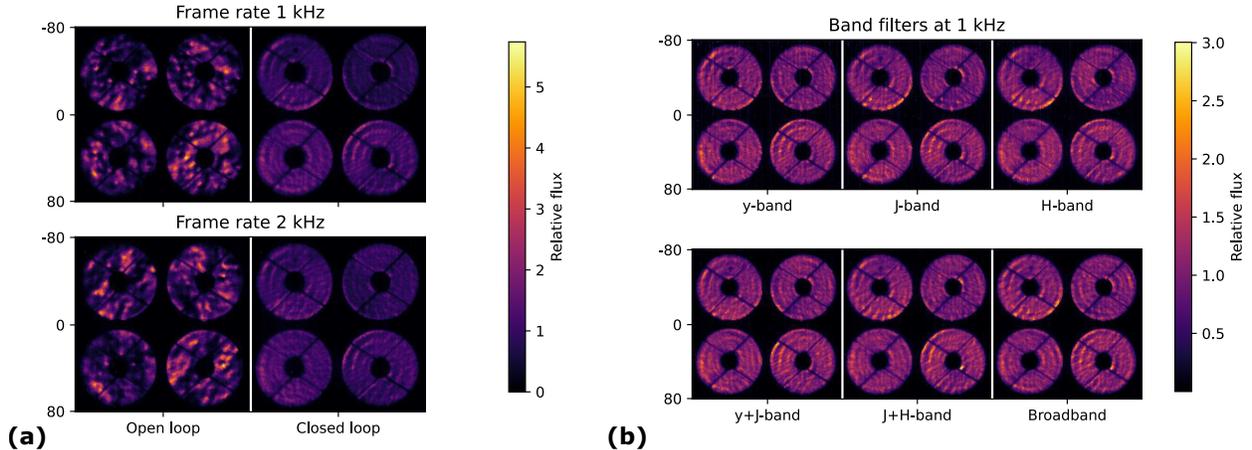} 
   \end{tabular}
   \end{center}
   \caption{(a) AO188 loop open and closed. (b) PyWFS images with the various filters. Average of ~30s of data for all cases.}
   \label{fig:pywfsonsky} 
   \end{figure} 

We took several telemetry datasets with the PyWFS, for various modes: with the AO188 loop open and closed, with the various filters available, as well as for different acquisition frequencies and modulation radii. In all cases, the PyWFS performed as expected. SCExAO telemetry (e.g. focal plane images in visible and NIR, visible PyWFS pseudo-open loop measurements) was acquired simultaneously, although some more work is needed to compare the results to the NIR PyWFS telemetry.

Figure~\ref{fig:pywfsonsky} (a) presents 30-s averages at 1~kHz and 2~kHz, with the AO188 loop open and closed, at maximum modulation. In open-loop, what we see is not really the turbulence, as it averages over 30~s, but the shape of the DM before the loop was opened. Though it is a good indication of the amount of signal we would get in open-loop in median seeing. 

Figure~\ref{fig:pywfsonsky} (b) shows 30-s averages for each filter. There is no noticeable differences between the different filters and the broadband mode, demonstrating the good achromaticity of the optics, as well as the limited chromatic effects of the double-roof pyramid design. A more dynamical analysis of the data is on-going, especially comparing them to the visible PyWFS telemetry, although the lack of response matrix for the NIR PyWFS is limiting us.

Despite not being able to close the loop with the NIR PyWFS, we can still deduce limits in stellar magnitudes by comparing the signal-to-noise ratio (SNR) per pixel obtained for the known magnitude with the minimum SNR per pixel needed to close the loop with the visible PyWFS inside SCExAO. Indeed, with SCExAO, we are able to close the loop up to $m_I = 11$ stars, giving a minimum SNR per pixel of $S_\textrm{min} = 0.12$.

For these measurements, we acquired data at 1~kHz on the A0V star HD 116160, which has a very flat magnitude of $m_\star=5.6$ over all the bands of interest. First, the average flux in the illuminated parts of the image $F_\textrm{ADU}$ is converted in electrons $F_\textrm{e}$, by multiplying with the detector gain $g_\textrm{det}=0.45$~e$^{-}$/ADU and the amplification gain $g_\textrm{amp}=121$ for this camera, such as
\begin{equation}
    F_\textrm{e} = F_\textrm{ADU}\times g_\textrm{det} \times g_\textrm{amp}.
\end{equation}

Then the SNR per pixel $S_\star$ is calculated by considering both the readout noise (RON) measured from the dark files $\sigma_\textrm{RON}=0.12$~e$^{-}$ and the photon noise estimated from the flux $\sigma_\textrm{ph}=\sqrt{F_\textrm{e}}$, such as
\begin{equation}
    S_\star = \frac{F_\textrm{e}}{\sqrt{\sigma_\textrm{ph}^2+\sigma_\textrm{RON}^2}}=\frac{F_\textrm{e}}{\sqrt{F_\textrm{e}+\sigma_\textrm{RON}^2}}.
\end{equation}
Finally, we find the magnitude limit at 1~kHz $m_\textrm{lim,1kHz}$ by comparing this measured SNR to the assumed minimum SNR $S_\textrm{min}$,
\begin{equation}
    m_\textrm{lim,1kHz} = m_\star+5\log{\frac{S_\star}{S_\textrm{min}}}.
\end{equation}
We can also interpolate this result for other frequencies $f$ by adding $2.5\log{t_f/t_\textrm{1kHz}}$ to the magnitude limit, where $t_f$ is the actual exposure time at that frequency. Finally, the magnitude limit for the IRCS dichroic can also be interpolated by adding $2.5\log{2}=0.75$, since the IRCS beamsplitter reflects 2 times more light than the SCExAO beamsplitter.

   \begin{figure} [t]
   \begin{center}
   \begin{tabular}{c}
   \includegraphics[width=0.97\textwidth]{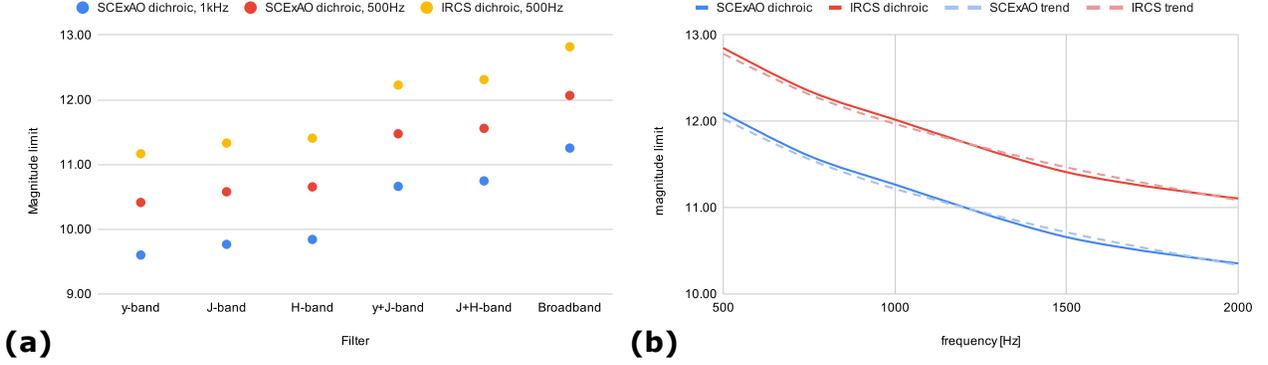} 
   \end{tabular}
   \end{center}
   \caption{(a) Estimated magnitude limit with the various filters, by matching the SNR per pixel to the visible PyWFS case. (b) Estimated magnitude limit in broadband for various loop frequencies.}
   \label{fig:pywfsmaglim} 
   \end{figure} 

Figure~\ref{fig:pywfsmaglim} (a) presents the calculated magnitude limits based on the datasets for each filter.  The band filters give similar results, with magnitude limits around 10.5 for the SCExAO dichroic at 500~Hz, and around 11.3 for the IRCS dichroic at the same frequency. With two bands, the magnitude limit increases by about one magnitude, and another half magnitude with all the light. Of course, this will depend heavily on the spectrum of the star: very red stars will provide light mostly in H-band for example. 

Figure~\ref{fig:pywfsmaglim} (b) presents the same measurements in broadband mode, for different frequencies, for the two beamsplitters. It follows the expected trend $2.5\log{t_f/t_\textrm{1kHz}}$.

We will be able to confirm these magnitude limits once we close the loop on-sky, hopefully before the end of 2022.

\subsection{First On-Sky Testing of the FPWFS Mode}
\label{sec:fpwfs}

   \begin{figure} [t]
   \begin{center}
   \begin{tabular}{c}
   \includegraphics[width=0.97\textwidth]{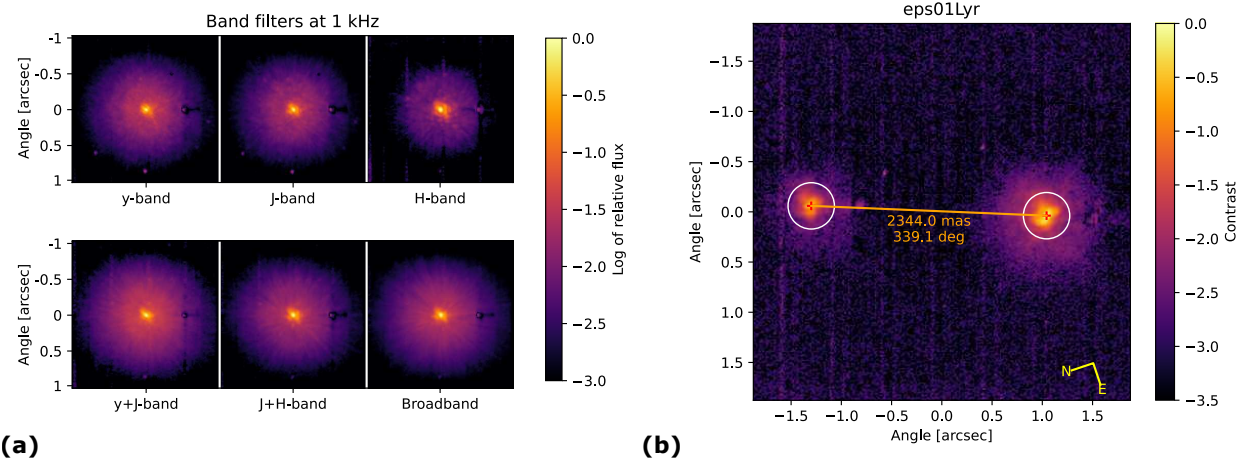} 
   \end{tabular}
   \end{center}
   \caption{(a) Focal plane images with various filters. (b) Astrometric calibration on the binary $\epsilon$01 Lyrae AB. The plate scale is estimated at 16~mas/pix.}
   \label{fig:fpwfsonsky} 
   \end{figure} 

The focal plane imaging mode was also tested on-sky, with the visible AO loop closed. In a first test, we took 30-s data sets in focus, for the various filters, on the A0V star HIP 65198 ($m_\star = 5.6$ over all bands). The data was taken at 2~kHz for all filters, but with different amplification gains to avoid saturation: $g_\textrm{amp}=64$ for y- and J-band, and $g_\textrm{16}$ for the other filters. The shift-and-averaged images are presented in Fig.~\ref{fig:fpwfsonsky} (a). The chromaticity of the Strehl ratio, PSF size and halo diameter is clearly visible in the band images. We can also notice that the speckles are washed out radially once more than one band is used. Note that the artifact seen on the right of each PSF is a numerical ghost coming from the detector itself, a copy of the image at 32 pixels (one readout column). It was subtracted simply by rescaling and shifting the image, although a better subtraction will be attempted in the future.

A known binary system, in this case $\epsilon$01 Lyrae AB (A3V+F0V $m_A = 5$ and $m_B=6$) was observed in H-band to calibrate the plate scale and orientation. An horizontal offset of 1" was added to AO188 to fit the binary inside the sub-window. The image is presented in Fig.~\ref{fig:fpwfsonsky} (b). A plate scale of 16.0~mas/pix provides the proper separation, although there is a discrepancy of $10^o$ with the last known measurement of this binary. Simultaneous measurements with CHARIS on another multiple star system will confirm if this discrepancy is instrumental or not. With this plate scale, the field-of-view of the 160x160 sub-window is 2.56"x2.56", while the full-frame image has a field-of-view of 5.1"x4.1", although internal vignetting kicks in for separations over 2". Thanks to the plate scale measurement, we will be able to calibrate the steering capability to close the loop on off-axis targets. It is also useful to calibrate the modulation radius for the PyWFS mode.

   \begin{figure} [b]
   \begin{center}
   \begin{tabular}{c}
   \includegraphics[width=0.97\textwidth]{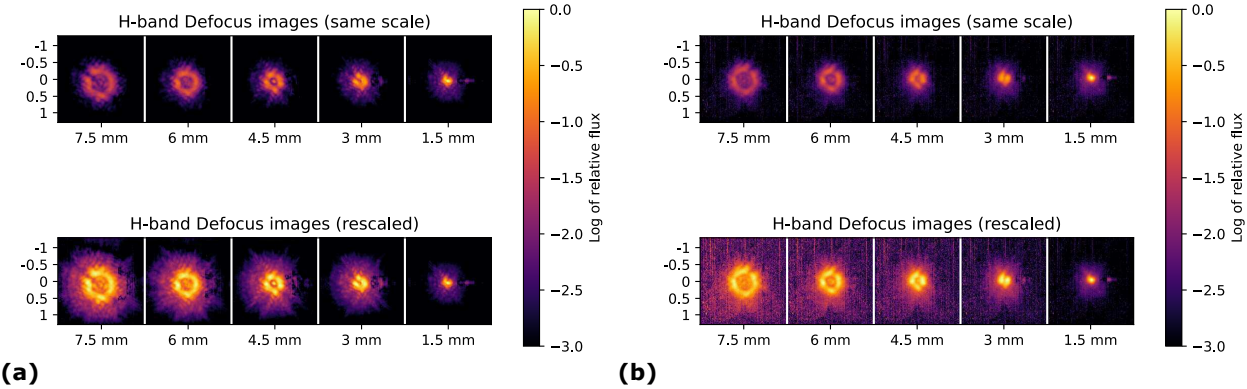} 
   \end{tabular}
   \end{center}
   \caption{H-band Focal plane images for various defocus values (translation of the diverging lens of the focusing optics) for two targets: (a) $\lambda$ Aquilae (mag 3.64) and (b) TYC 514025931 (mag 9.7).}
   \label{fig:fpwfsdefocus} 
   \end{figure}

The defocus necessary for the FPWFS mode can be achieved by changing the separation between the converging and diverging lenses of the focusing optics: the diverging lens is on a motorized translation stage. We took 30-s data sets with various defocus positions, with the H-band filter, for two stars of different magnitudes: $\lambda$ Aquilae ($m_\star=3.64$) and TYC 514025931 ($m_\star=9.68$).  The largest defocus can spread the PSF over $\sim60$~pixels in diameter. The resulting images are presented in Fig.~\ref{fig:fpwfsdefocus}, by keeping the same scale for all the defocus values (top), and by normalizing each of them with the maximum value (bottom). 
  
 \begin{figure} [t]
   \begin{center}
   \begin{tabular}{c}
   \includegraphics[width=0.97\textwidth]{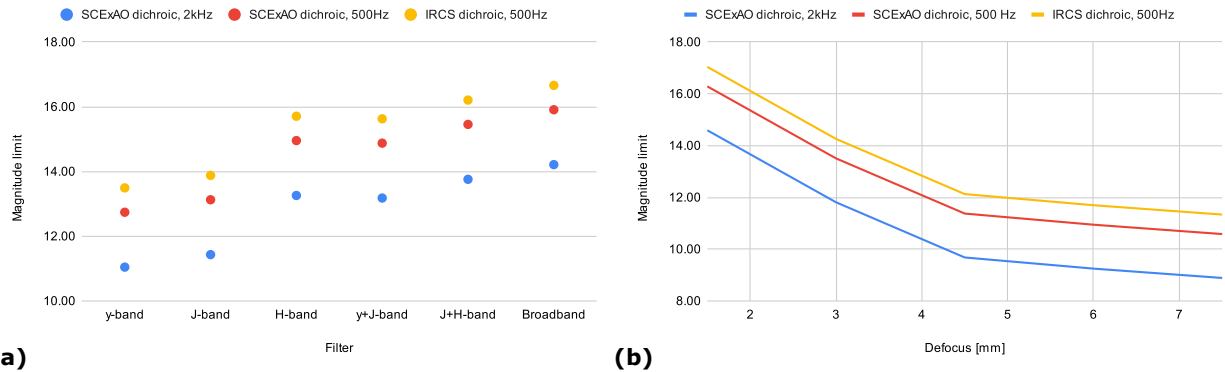} 
   \end{tabular}
   \end{center}
   \caption{(a) Estimated magnitude limit for various filters. (b) Estimated magnitude limit (H-band) for various defocus values.}
   \label{fig:fpwfsmaglim} 
   \end{figure} 

Similarly to the PyWFS mode, we tried to estimate the magnitude limits by calculating the SNR of the brightest pixel and in this case considered a minimum SNR of 0.6 , i.e. 5 times the minimum SNR per pixel of the PyWFS mode. We only have in-focus images for all the filters, so the estimations presented in Fig.~\ref{fig:fpwfsmaglim} (a) are probably optimistic, since a defocus is needed for wavefront control. Finally, Fig.~\ref{fig:fpwfsmaglim} (b) presents the magnitude limits for different defocus values in H-band, based on the brightest target. More tests are needed to see if those are realistic or not, but we would expect to be able to reach fainter targets than the PyWFS modes, although we would correct fewer modes.

The results presented in Sec.~\ref{sec:nirwfs} show the success of Phase Ia, although the steering capacity still needs to be demonstrated on-sky. The full demonstration of the NIR WFS will have to wait for Phase Ib, with the integration of the new 64x64 DM, to be able to close the loop. 

\section{PHASE Ib: TESTING OF THE NEW ALPAO DM IN THE LAB BEFORE INTEGRATION}
\label{sec:dm}

In Phase Ib, we will replace the current bimorph DM with the new ALPAO 64x64 DM. The delivery of the ALPAO DM was delayed to the end of March 2022, and for contracting reasons it had to be delivered despite not being fully ready: the flatness was not up to specs. We will send it back in August 2022 so they can improve the flatness, and get it back for the engineering test in November. 

Despite this setback, the first delivery let us test the custom manual pitch/yaw mount designed to lift the DM up to the 250-mm beam height, and test the manual alignment of the DM. We also did some extensive work to send commands to the electronics, in a way compatible with the CACAO framework that will be used by AO3000.

   \begin{figure} [t]
   \begin{center}
   \begin{tabular}{c}
   \includegraphics[width=0.7\textwidth]{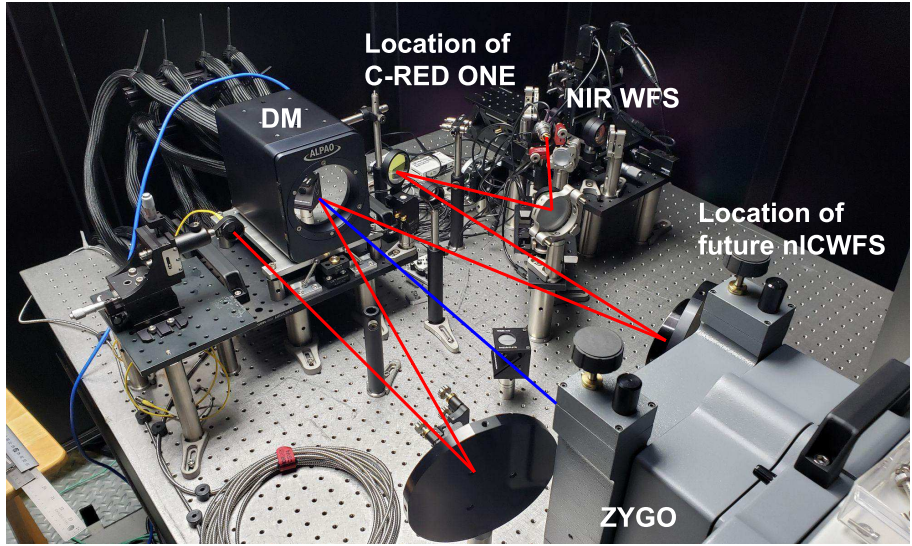} 
   \end{tabular}
   \end{center}
   \caption{Laboratory testbench of the ALPAO DM (top left) and the NIR WFS (top right). A Zygo is facing the DM (bottom right) to simultaneously take calibrated wavefront measurements. In the future, we will add the non-linear visible curvature WFS on the same bench.}
   \label{fig:dmtesting} 
   \end{figure} 

Scheduling the replacement of the DM is a challenge, since we want to limit as much as possible the impact on science observations. That is why we will keep the ability to swap efficiently between the original DM and the new DM. So when the new DM is not in place inside AO188, it will come down the Subaru offices in Hilo, Hawai`i, where we are building a test setup. This testbench includes a polychromatic supercontinuum source, the DM, the NIR WFS (also removed from AO188 after the on-sky tests), and a Zygo interferometer facing the DM. The new visible non-linear curvature WFS will also be included there in the future (see Fig.~\ref{fig:dmtesting}). Having both the NIR WFS and the Zygo at the same time will allow us to test close-loop operations with the PyWFS or the FPWFS mode, and have an absolute measurement of the shape of the DM at the same time.

The second C-RED ONE camera was received at the end of June, so we will be able to use it on the testbed for the NIR WFS, while the first one will stay at the summit for the FPDI mode of SCExAO.

The testbed uses mostly off-the-shelf opto-mechanical parts: an off-axis parabola (OAP, 100~mm diameter, 500~mm focal length) collimates the light on the DM, and an identical OAP focuses the light again. The we use a diverging doublet to match the focal ratio of 13.9 at the output of AO188. But since achromatic off-the-shelf NIR lenses are impossible to find, the lens is only achromatic up to $\sim1$~\mum, and displays some large chromatic shift at higher wavelengths. Therefore the current testbed could only be used in y-band. This should be fine to perform close-loop tests, although we can potentially swap the lens with a custom achromatic doublet in NIR, similar to the ones already used in the NIR WFS.

\section{CONCLUSION AND PERSPECTIVE}

To fully exploit the high-contrast imaging potential of the Subaru Telescope, the facility adaptive optics AO188 will receive well deserved upgrades. Despite some delays, we are making great progress towards upgrading AO188 into AO3000, a XAO-level first stage adaptive optics with more than 3000 actuators in the pupil. The upgrades will happen in stages, to minimize the impact on current science observations. Once everything is in place, AO3000 will feed multiple instruments thanks to a beam switcher, and even allow for simultaneous observations with two instruments.

In a first step, We successfully installed and tested on-sky for the first time a NIR WFS using a C-RED ONE camera. With the NIR WFS, we achieved the goal of adding two wavefront sensing capabilities (a pyramid WFS mode and a focal plane WFS mode) to AO188, without compromising any of its capabilities. The first on-sky tests, despite being brief and only in open-loop, taught us a lot about the future capabilities of this WFS. The PyWFS should be able to close the loop down to magnitudes 10 to 13, depending on the spectral type of the star and the mode used. The FPWFS should be able to reach even fainter stars, although more lab and on-sky work is needed to confirm these numbers. 

We also received the new DM that will power AO3000, and built a testbed in the Subaru offices in Hilo, Hawai`i, where we will be able to perform close-loop tests using the NIR WFS and a future visible non-linear CWFS, while at the same time monitoring the shape of the DM surface with a Zygo interferometer. These tests were delayed due to some issues with the DM itself, which should be resolved before the end of the year. If all goes well, we will do a first on-sky test of the DM at the end of November 2022. 

The next few years is truly going to be exciting for high-contrast imaging at the Subaru Telescope, as we will be able to reach deeper and more stable contrasts, around much redder targets. The technologies tested now will also be essential for the next generation of telescopes like TMT.

\acknowledgments 

The development of SCExAO \& AO3000 were supported by the Japan Society for the Promotion of Science (Grant-in-Aid for Research \#23340051, \#26220704, \#23103002, \#19H00703 \& \#19H00695), the Astrobiology Center of the National Institutes of Natural Sciences, Japan, the Mt Cuba Foundation and the director's contingency fund at Subaru Telescope.
The authors wish to recognize and acknowledge the very significant cultural role and reverence that the summit of Maunakea has always had within the Hawaiian community.
We are most fortunate to have the opportunity to conduct observations from this mountain.
KA acknowledges funding from the Heising-Simons foundation.
VD acknowledges support from NASA funding (Grant \#80NSSC19K0336).

\bibliography{report} 

\begin{thebibliography}{10}

\bibitem{Takami2006}
{Takami}, H., {Colley}, S., {Dinkins}, M., {Eldred}, M., {Guyon}, O., {Golota},
  T., {Hattori}, M., {Hayano}, Y., {Ito}, M., {Iye}, M., {Oya}, S., {Saito},
  Y., and {Watanabe}, M., ``{Status of Subaru laser guide star AO system},'' in
  [{\em Society of Photo-Optical Instrumentation Engineers (SPIE) Conference
  Series}{\nolinebreak\hspace{0.1em}]},  {Ellerbroek}, B.~L. and {Bonaccini
  Calia}, D., eds., {\em Society of Photo-Optical Instrumentation Engineers
  (SPIE) Conference Series} {\bf 6272},  62720C (June 2006).

\bibitem{Hayano2006}
{Hayano}, Y., {Saito}, Y., {Ito}, M., {Saito}, N., {Kato}, M., {Akagawa}, K.,
  {Takazawa}, A., {Colley}, S.~A., {Dinkins}, M.~C., {Eldred}, M., {Golota},
  T.~I., {Guyon}, O., {Hattori}, M., {Oya}, S., {Watanabe}, M., {Takami}, H.,
  {Wada}, S., and {Iye}, M., ``{The laser guide star facility for Subaru
  Telescope},'' in [{\em Society of Photo-Optical Instrumentation Engineers
  (SPIE) Conference Series}{\nolinebreak\hspace{0.1em}]},  {Ellerbroek}, B.~L.
  and {Bonaccini Calia}, D., eds., {\em Society of Photo-Optical
  Instrumentation Engineers (SPIE) Conference Series} {\bf 6272},  627247 (June
  2006).

\bibitem{Minowa2010}
{Minowa}, Y., {Hayano}, Y., {Oya}, S., {Watanabe}, M., {Hattori}, M., {Guyon},
  O., {Egner}, S., {Saito}, Y., {Ito}, M., {Takami}, H., {Garrel}, V.,
  {Colley}, S., {Golota}, T., and {Iye}, M., ``{Performance of Subaru adaptive
  optics system AO188},'' in [{\em Adaptive Optics Systems
  II}{\nolinebreak\hspace{0.1em}]},  {\em Proc.\ Soc.\ Photo-Opt.\ Instrum.\
  Eng.} {\bf 7736},  77363N (July 2010).

\bibitem{Jovanovic2015}
{Jovanovic}, N., {Martinache}, F., {Guyon}, O., {Clergeon}, C., {Singh}, G.,
  {Kudo}, T., {Garrel}, V., {Newman}, K., {Doughty}, D., {Lozi}, J., {Males},
  J., {Minowa}, Y., {Hayano}, Y., {Takato}, N., {Morino}, J., {Kuhn}, J.,
  {Serabyn}, E., {Norris}, B., {Tuthill}, P., {Schworer}, G., {Stewart}, P.,
  {Close}, L., {Huby}, E., {Perrin}, G., {Lacour}, S., {Gauchet}, L.,
  {Vievard}, S., {Murakami}, N., {Oshiyama}, F., {Baba}, N., {Matsuo}, T.,
  {Nishikawa}, J., {Tamura}, M., {Lai}, O., {Marchis}, F., {Duchene}, G.,
  {Kotani}, T., and {Woillez}, J., ``{The Subaru Coronagraphic Extreme Adaptive
  Optics System: Enabling High-Contrast Imaging on Solar-System Scales},'' {\em
  Pub.\ Astron.\ Soc.\ Pacific}~{\bf 127},  890 (Sep 2015).

\bibitem{Lozi2018}
{Lozi}, J., {Guyon}, O., {Jovanovic}, N., {Goebel}, S., {Pathak}, P., {Skaf},
  N., {Sahoo}, A., {Norris}, B., {Martinache}, F., {N'Diaye}, M., {Mazin}, B.,
  {Walter}, A.~B., {Tuthill}, P., {Kudo}, T., {Kawahara}, H., {Kotani}, T.,
  {Ireland}, M., {Cvetojevic}, N., {Huby}, E., {Lacour}, S., {Vievard}, S.,
  {Groff}, T.~D., {Chilcote}, J.~K., {Kasdin}, J., {Knight}, J., {Snik}, F.,
  {Doelman}, D., {Minowa}, Y., {Clergeon}, C., {Takato}, N., {Tamura}, M.,
  {Currie}, T., {Takami}, H., and {Hayashi}, M., ``{SCExAO, an instrument with
  a dual purpose: perform cutting-edge science and develop new technologies},''
  in [{\em Proc.\ Soc.\ Photo-Opt.\ Instrum.\
  Eng.}{\nolinebreak\hspace{0.1em}]},  {\em Society of Photo-Optical
  Instrumentation Engineers (SPIE) Conference Series} {\bf 10703},  1070359
  (Jul 2018).

\bibitem{Kotani2018}
{Kotani}, T., {Tamura}, M., {Nishikawa}, J., {Ueda}, A., {Kuzuhara}, M.,
  {Omiya}, M., {Hashimoto}, J., {Ishizuka}, M., {Hirano}, T., {Suto}, H.,
  {Kurokawa}, T., {Kokubo}, T., {Mori}, T., {Tanaka}, Y., {Kashiwagi}, K.,
  {Konishi}, M., {Kudo}, T., {Sato}, B., {Jacobson}, S., {Hodapp}, K.~W.,
  {Hall}, D.~B., {Aoki}, W., {Usuda}, T., {Nishiyama}, S., {Nakajima}, T.,
  {Ikeda}, Y., {Yamamuro}, T., {Morino}, J.-I., {Baba}, H., {Hosokawa}, K.,
  {Ishikawa}, H., {Narita}, N., {Kokubo}, E., {Hayano}, Y., {Izumiura}, H.,
  {Kambe}, E., {Kusakabe}, N., {Kwon}, J., {Ikoma}, M., {Hori}, Y., {Genda},
  H., {Fukui}, A., {Fujii}, Y., {Kawahara}, H., {Olivier}, G., {Jovanovic}, N.,
  {Harakawa}, H., {Hayashi}, M., {Hidai}, M., {Machida}, M., {Matsuo}, T.,
  {Nagata}, T., {Ogihara}, M., {Takami}, H., {Takato}, N., {Terada}, H., and
  {Oh}, D., ``{The infrared Doppler (IRD) instrument for the Subaru telescope:
  instrument description and commissioning results},'' in [{\em Proc.\ Soc.\
  Photo-Opt.\ Instrum.\ Eng.}{\nolinebreak\hspace{0.1em}]},  {Evans}, C.~J.,
  {Simard}, L., and {Takami}, H., eds., {\em Society of Photo-Optical
  Instrumentation Engineers (SPIE) Conference Series} {\bf 10702},  1070211
  (July 2018).

\bibitem{Ono2020}
{Ono}, Y.~H., {Minowa}, Y., {Guyon}, O., {Clergeon}, C.~S., {Mieda}, E.,
  {Lozi}, J., {Hattori}, T., and {Akiyama}, M., ``{Overview of AO activities at
  Subaru Telescope},'' in [{\em Society of Photo-Optical Instrumentation
  Engineers (SPIE) Conference Series}{\nolinebreak\hspace{0.1em}]},  {\em
  Society of Photo-Optical Instrumentation Engineers (SPIE) Conference Series}
  {\bf 11448},  114480K (Dec. 2020).

\bibitem{Vidal2019}
{Vidal}, F., {Raffard}, J., {Gendron}, E., {Thijs}, S., {Lapeyr{\`e}re}, V.,
  {Buey}, T., {Cl{\'e}net}, Y., {Gratadour}, D., {Jagourel}, P., {Sevin}, A.,
  {Ferreira}, F., {Chemla}, F., and {Mahiou}, P., ``{Tests and
  characterisations of the ALPAO 64{\texttimes}64 deformable mirror, the
  MICADO-MAORY SCAO AIT facility},'' in [{\em Proceedings of the AO4ELT6
  conference}{\nolinebreak\hspace{0.1em}]},   E4 (Nov. 2019).

\bibitem{Feldt2006a}
{Feldt}, M., {Peter}, D., {Hippler}, S., {Henning}, T., {Aceituno}, J., and
  {Goto}, M., ``{PYRAMIR: first on-sky results from an infrared pyramid
  wavefront sensor},'' in [{\em Society of Photo-Optical Instrumentation
  Engineers (SPIE) Conference Series}{\nolinebreak\hspace{0.1em}]},
  {Ellerbroek}, B.~L. and {Bonaccini Calia}, D., eds., {\em Society of
  Photo-Optical Instrumentation Engineers (SPIE) Conference Series} {\bf 6272},
   627218 (June 2006).

\bibitem{Feldt2006b}
{Feldt}, M., {Hayano}, Y., {Takami}, H., {Usuda}, T., {Watanabe}, M., {Iye},
  M., {Goto}, M., {Bizenberger}, P., {Egner}, S., and {Peter}, D., ``{SUPY: an
  infrared pyramid wavefront sensor for Subaru},'' in [{\em Society of
  Photo-Optical Instrumentation Engineers (SPIE) Conference
  Series}{\nolinebreak\hspace{0.1em}]},  {Ellerbroek}, B.~L. and {Bonaccini
  Calia}, D., eds., {\em Society of Photo-Optical Instrumentation Engineers
  (SPIE) Conference Series} {\bf 6272},  62722A (June 2006).

\bibitem{Atkinson2018}
{Atkinson}, D., {Hall}, D., {Goebel}, S., {Jacobson}, S., and {Baker}, I.,
  ``{Observatory deployment and characterization of SAPHIRA HgCdTe APD
  arrays},'' in [{\em High Energy, Optical, and Infrared Detectors for
  Astronomy VIII}{\nolinebreak\hspace{0.1em}]},  {Holland}, A.~D. and
  {Beletic}, J., eds., {\em Society of Photo-Optical Instrumentation Engineers
  (SPIE) Conference Series} {\bf 10709},  107091H (July 2018).

\bibitem{Lozi2019}
{Lozi}, J., {Jovanovic}, N., {Guyon}, O., {Chun}, M., {Jacobson}, S., {Goebel},
  S., and {Martinache}, F., ``{Visible and Near Infrared Laboratory
  Demonstration of a Simplified Pyramid Wavefront Sensor},'' {\em Pub.\
  Astron.\ Soc.\ Pacific}~{\bf 131},  044503 (Apr 2019).

\bibitem{Lozi2020}
{Lozi}, J., {Guyon}, O., {Kudo}, T., {Zhang}, J., {Jovanovic}, N., {Norris},
  B., {Martinod}, M.-A., {Groff}, T.~D., {Chilcote}, J., {Tamura}, M., {Bos},
  S., {Snik}, F., {Vievard}, S., {Sahoo}, A., {Deo}, V., {Martinache}, F., and
  {Kasdin}, J., ``{New NIR spectro-polarimetric modes for the SCExAO
  instrument},'' in [{\em Society of Photo-Optical Instrumentation Engineers
  (SPIE) Conference Series}{\nolinebreak\hspace{0.1em}]},  {\em Society of
  Photo-Optical Instrumentation Engineers (SPIE) Conference Series} {\bf
  11448},  114487C (Dec. 2020).

\bibitem{Singh2015}
{Singh}, G., {Lozi}, J., {Guyon}, O., {Baudoz}, P., {Jovanovic}, N.,
  {Martinache}, F., {Kudo}, T., {Serabyn}, E., and {Kuhn}, J., ``{On-Sky
  Demonstration of Low-Order Wavefront Sensing and Control with Focal Plane
  Phase Mask Coronagraphs},'' {\em Pub.\ Astron.\ Soc.\ Pacific}~{\bf 127},
  857--869 (Oct. 2015).

\bibitem{Vievard2020}
{Vievard}, S., {Bos}, S.~P., {Cassaing}, F., {Currie}, T., {Deo}, V., {Guyon},
  O., {Jovanovic}, N., {Keller}, C.~U., {Lamb}, M., {Lopez}, C., {Lozi}, J.,
  {Martinache}, F., {Miller}, K., {Montmerle-Bonnefois}, A., {Mugnier}, L.~M.,
  {N'Diaye}, M., {Norris}, B., {Sahoo}, A., {Sauvage}, J.~F., {Skaf}, N.,
  {Snik}, F., {Wilby}, M.~J., and {Wong}, A., ``{Focal plane wavefront sensing
  on SUBARU/SCExAO},'' in [{\em Society of Photo-Optical Instrumentation
  Engineers (SPIE) Conference Series}{\nolinebreak\hspace{0.1em}]},  {\em
  Society of Photo-Optical Instrumentation Engineers (SPIE) Conference Series}
  {\bf 11448},  114486D (Dec. 2020).

\end{thebibliography}
\bibliographystyle{spiebib} 

\end{document}